\documentstyle[12pt]{article}

\newtheorem{theorem}{Theorem}
\newtheorem{lemma}{Lemma}
\newtheorem{corollary}{Corollary}

\begin{document}

\newcommand{\lap}{\bigtriangleup}
\def\be{\begin{equation}}
\def\ee{\end{equation}}
\def\bea{\begin{eqnarray}}
\def\eea{\end{eqnarray}}
\def\beas{\begin{eqnarray*}}
\def\eeas{\end{eqnarray*}}
\def\n#1{\vert #1 \vert}
\def\nn#1{{\Vert #1 \Vert}}
\def\R{\mathrm{ I\kern-.1567em R}}
\def\N{\mathrm{ I\kern-.1567em N}}
\def\D{{\cal D}}
\def\F{{\cal F}}
\def\M{{\cal M}}
\def\supp{\mbox{\rm supp}}
 
\title{Stable Steady States in Stellar Dynamics}
\author{Yan Guo\\
        Lefschetz Center for Dynamical Systems\\
        Division of Applied Mathematics\\
        Brown University, Providence, RI 02912\\
        \ \\ 
        Gerhard Rein\\
        Mathematisches Institut der Universit\"at M\"unchen\\
        Theresienstr.\ 39, 80333 M\"unchen, Germany}         
\date{}
\maketitle
\begin{abstract}
We prove the existence and nonlinear stability of steady states of 
the Vlasov-Poisson system in the stellar dynamics case. The steady states 
are obtained as minimizers of an energy-Casimir functional from which 
fact their dynamical stability is deduced. 
The analysis applies to some of the well-known
polytropic steady states, but it also considerably extends the class
of known steady states.
\end{abstract}

\section{Introduction}
\setcounter{equation}{0}

A galaxy or a globular cluster can be modelled as an 
ensemble of particles, i.~e., stars, which interact only 
by the gravitational field which they create collectively,
collisions among the particles being sufficiently rare to
be neglected. In a Newtonian setting the time evolution
of such an ensemble is governed by the Vlasov-Poisson system: 
\be \label{vlasov}
\partial_t f + v \cdot \nabla _x f - \nabla _x U \cdot 
\nabla _v f = 0,
\ee
\be \label{poisson} 
\lap U = 4 \pi\, \rho, 
\ee
\be \label{rho}
\rho(t,x)= \int f(t,x,v)dv .
\ee
Here $f = f(t,x,v)\geq 0$ denotes the density of the particles 
in phase space, $t \in \R$ denotes time, $x, v \in \R^3$ denote 
position and velocity respectively, $\rho$ is the spatial mass 
density, and $U$ the gravitational potential.

In the present paper we are interested in the existence and
stability of steady states of this system. Up to now, steady states
for the Vlasov-Poisson system have been contructed in the following
way: If $U$ is time independent, the particle energy
\be \label{parten}
E=\frac{1}{2}|v|^2 + U (x) 
\ee
is conserved along characteristics of (\ref{vlasov}), if $U$ in 
addition is spherically symmetric the same is true for the modulus
of angular momentum
\be \label{angmom}
L=|x \times v|^2 = |v|^2|x|^2-(x \cdot v)^2.
\ee
Thus the ansatz
\[
f(x,v)=\phi(E,L)
\]
satisfies the Vlasov equation and reduces the time independent 
Vlasov-Poisson system to the semilinear Poisson equation
\be \label{slpoisson}
\lap U = 4 \pi h(r,U)
\ee
where 
\[
h(r,u) = \int \phi\left(\frac{1}{2}|v|^2 + u,L\right)\, dv,\ r =\n{x}.
\]
In \cite{BFH} this procedure was carried out for the so-called polytropes
\be \label{poly}
\phi(E,L)=(E_0-E)_+^\mu L^k;
\ee
$(\cdot)_+$ denotes the positive part. 
The crucial issue there is to show that a solution
of (\ref{slpoisson})---once its existence is established---leads
to a steady state with finite mass and compact support,\ cf.\
also \cite{BP}. The approach to this problem used so far is
highly dependent on the particular form of $\phi$.

In the present paper we construct steady states as minimizers of
an appropriately defined energy-Casimir functional. Given a function
$Q=Q(f,L) \geq 0,\ f,\ L \geq 0$, we define
\be \label{j}
J(f) :=
\int\!\!\int Q(f,L)\,dv\,dx +{1\over 2} \int\!\!\int |v|^2 f\,dv\,dx
\ee
and
\be \label{cd}
\D(f) := 
J(f) - \frac{1}{8\pi} \int |\nabla U_f|^2\,dx.
\ee
Here $f=f(x,v)$ is taken from some appropriate set $\F_M$
of functions which in particular have total mass equal to a 
prescribed constant $M$ and are spherically symmetric, 
and $U_f$ denotes the potential induced by $f$ with
boundary value $0$ at spatial infinity.
To obtain steady states of the Vlasov-Poisson system as minimizers
of the energy-Casimir functional $\D$ has the following
advantages: 
The approach does not rely on a particular form of ansatz like (\ref{poly})
so that a broader class of steady states is obtained. The finite
mass condition is built into the set of functions $\F_M$
and does not pose an extra problem, and we obtain an explicit bound
on the spatial support of the minimizers. Finally and most importantly,
the obtained steady states are stable. Throughout we consider only 
spherically symmetric functions $f$ so that the stability holds with
respect to spherically symmetric perturbations.

We now describe in more detail how the paper proceeds. In the next section
we state the assumptions for the function $Q$ which determines our
energy-Casimir functional, and prove some preliminary results, in
particular a lower bound of $\D$ on $\F_M$. 
The main difficulties in finding a minimizer of $\D$
arise from the fact that $\D$ is neither positive definite 
nor convex, and from the lack of compactness: Along
a minimizing sequence some mass could escape to infinity. This is 
impossible due to two crucial observations which are established 
in the third section. The first one is 
based on a scaling argument and asserts 
that $\D_M=\inf_{\F_M} \D (f) <0$ and 
\[
\D_{M_1}\ge \left({{M_1}\over {M_2}}\right)^{1+\alpha}
\D_{M_2}
\]
for all positive $M_1\le M_2$ and some $\alpha>0$. 
The second one is based 
on a splitting argument. We split the physical space into
$B_R=\{|x|\le R\}$ and its complement. From the above 
scaling identity we obtain the estimate
\[
\D (f)-\D_M \geq \left(-{{C_\alpha  \D_M}\over{M^2}}
-{1\over {R}}\right) m_f(R)\,(M-m_f(R)).
\]
Here $C_\alpha>0$ is a constant and
$m_f(R)=\int_{B_R}\int  f$. This implies that along
any minimizing sequence the mass has to 
concentrate in a fixed ball. In the fourth section we use this to
show that a minimizer of $\D$ over $\F_M$ exists, and we prove that
every such minimizer is a steady state of the Vlasov-Poisson system.
In particular we show that
the gravitational field $\nabla U_f$ converges strongly in
$L^2(\R^3)$ along any minimizing sequence.
Dynamical stability of such a minimizer $f_0$ then follows easily
from the fact that $\D$ is conserved along spherically symmetric
solutions of the Vlasov-Poisson system. 
To measure the distance of a perturbation 
from $f_0$ we use the quantity
\[
\int\!\!\int \Bigl[Q(f,L)-Q(f_0, L)+(E-E_0)(f-f_0)\Bigr]\,dv\,dx +
\frac{1}{8\pi} \|\nabla U_f-\nabla U_{f_0} \|_2^2;
\]
the first term turns out to be nonnegative.
Similar constructions have been 
used in the previous study of stability in collisionless 
plasmas by the one of the authors \cite{G1,G2}. 
A delicate point arises from the question whether the 
minimizer is unique in the set $\F_M$.
This can be shown in the case of the polytropes.
For the general case we had to leave this question open,
which results in the fact that we then obtain the stability
only with respect to the whole set of minimizers.
 
We conclude this introduction with some references to the 
literature. The existence of global classical solutions has been 
shown in \cite{P} as well as in \cite{H,LP,S}. The existence
of steady states for the case of the polytropic ansatz
was investigated in \cite{BFH} and \cite{BP}. There have
been many contributions to the stability problem in the astrophysics
literature; we refer to the monograph \cite{FP}. As far as
mathematically rigorous results are concerned, we mention
\cite{Wo}, where the stability of the polytropes is investigated
using a variational approach for a reduced energy-Casimir
functional defined on the space of mass functions
$m(r)=4 \pi \int_0^r s^2 \rho(s)\, ds$, and an investigation
of linearized stability in \cite{BMR}. For the plasma physics case,
where the sign in the Poisson equation (\ref{poisson}) is
reversed, the stability problem is much easier and better
understood. We refer to \cite{BRV,GS1,GS2,R2}.

\section{Preliminaries; a Lower Bound for $\D$}
\setcounter{equation}{0}

We first state the assumptions on $Q$ which are needed in the following:

\smallskip
\noindent
{\bf Assumptions on Q}: For $Q \in C^{1,0}([0,\infty[ \times [0,\infty[)\cap 
C^{2,0} (]0,\infty[\times [0,\infty[)$, $Q \geq 0$, and
constants $C_1,\ldots,C_4 >0$, $F_0 >0$, and 
$0 < \mu_1,\,\mu_2,\,\mu_3 < 3/2$
consider the following assumptions:
\begin{itemize}
\item[(Q1)]
$Q(f,L) \geq C_1 f^{1+1/{\mu_1}},\ f \geq F_0,\
L \geq 0$,
\item[(Q2)]
$Q(f,L) \leq C_2 f^{1+1/{\mu_2}},\ 0 \leq f \leq F_0,\
L \geq 0$,
\item[(Q3)]
$Q(\lambda f,L) \geq \lambda^{1+1/{\mu_3}} Q(f,L),\ f \geq 0,\ 
0 \leq \lambda \leq 1,\ L \geq 0$, and $Q(f,\cdot)$
is decreasing for all $f\geq 0$ if $\mu_3 < 1/2$,
\item[(Q4)]
$\partial_f^2 Q(f,L) > 0,\ f > 0,\ L \geq 0$, and 
$\partial_f Q(0,L) = 0,\ L \geq 0$,
\item[(Q5)]
$C_3 \partial_f^2 Q(f,L) \leq \partial_f^2 Q(\lambda f,L) 
\leq C_4 \partial_f^2 Q(f,L)$ for $f > 0,\ L \geq 0$ and $\lambda$
in some neighborhood of 1.
\end{itemize}
The above assumptions imply that for fixed $L \geq 0$ the
function $\partial_f Q(\cdot,L)$ is strictly
increasing with range $[0,\infty[$, and we denote its inverse
by $q(\cdot,L)$, i.~e., 
\be \label{qdef}
\partial_f Q(q(e,L),L) = e,\ e\geq 0,\ L\geq 0;
\ee
we extend $q(\cdot,L)$ by $q(e,L)=0,\ e < 0$.

\smallskip
\noindent
{\bf Remark:} The steady states obtained later will be of the
form 
\[
f_0 (x,v) = q(E_0 -E,L) 
\]
with some $E_0<0$ and $E$ and $L$ as defined in 
(\ref{parten}) and (\ref{angmom}) respectively.
If we take $Q(f,L) = f^{1+1/\mu},\ f\geq 0$, this leads to the
polytropic ansatz (\ref{poly}) with $k=0$, and such a $Q$ satisfies
the assumptions above if $0<\mu<3/2$. If we take
\be \label{nopol}
Q(f,L) = f^{1+1/{\mu_1}}\psi_1(L) +  f^{1+1/{\mu_2}}\psi_2(L)
\ee
with $0 < \mu_1,\,\mu_2 < 3/2$ and continuous functions
$\psi_1,\,\psi_2$ with $0< c \leq \psi_1(L),\,\psi_2(L) \leq C < \infty,
\ L\geq 0$, both decreasing if $\mu_1<1/2$ or $\mu_2<1/2$,
then again the above assumptions hold, but $q$ is clearly not
of polytropic form.

\smallskip

We will minimize the energy-Casimir functional $\D$ over the set
\bea \label{spacedef}
\F_M := \Bigl\{ f \in L^1(\R^6) 
&\mid&
f \geq 0,\ \int\!\!\int f dv\,dx = M, \nonumber\\
&&
J(f) < \infty,\ \mbox{and}\ f\ \mbox{is spherically symmetric}\Bigr\},
\eea
where $M>0$ is prescribed. Here spherical symmetry means that
\[
f(Ax,Av)=f(x,v),\ x,v \in \R^3,\ A \in \mbox{\rm SO}(3).
\]

The aim of the present section is to establish a lower bound
for $\D$ of a form that will imply the boundedness of $J$
along any minimizing sequence. To this end we first establish two
technical lemmas:

\begin{lemma} \label{rhoest}
Let $n = 3/2+\mu$ and $\mu>0$. Then there exists a constant
$C>0$ such that for all measurable $f\geq 0$,
\[
\int \rho_f^{1+1/n} dx\le C \left(\int\!\!\int f^{1+1/\mu} dv\,dx +
\int\!\!\int |v|^2 f\, dv\,dx \right)
\]
where
\[
\rho_f(x) = \int f(x,v)\, dv .
\]
\end{lemma}

\noindent
{\bf Proof}.  For any $R>0$,
\beas
\rho_f (x) 
&=&
\int_{\n{v} \leq R} f (x,v)\,dv + \int_{\n{v} \geq R} f (x,v)\,dv\\
&\leq&
C R^{3/(1+\mu)}\left(\int f^{1+1/\mu} dv \right)^{\mu/(\mu+1)}
+ \frac{1}{R^2} \int |v|^2 f\, dv
\eeas
by H\"older's inequality.
Optimizing the right hand side with respect to $R$, taking both sides of
the inequality to the power $1+1/n$ and integrating with respect to $x$
yields
\beas
\int \rho_f^{1+1/n} dx
&\leq&
C \int \left[\left(\int f^{1+1/\mu} dv \right)^{2 \mu/(5+2\mu)}
\left(\int |v|^2 f\, dv\right)^{3/(5+2\mu)}\right]^{1+1/n} dx \\
&\leq&
C \int \left(\int f^{1+1/\mu} dv +
\int |v|^2 f\, dv\right)^{(1+1/n)(3+2\mu)/(5+2\mu)} dx,
\eeas
and since $1+1/n=(5+2\mu)/(3+2\mu)$ this is the assertion.\hfill {\it QED}

\begin{lemma} \label{uest}
Let $\rho \in L^{1+1/n}(\R^3)$ be nonnegative and spherically symmetric
with $\int \rho = M$ and $1\leq n<3$.
Define 
\[
U_\rho := - \frac{1}{\n{\cdot}} \ast \rho .
\]
\begin{itemize}
\item[(a)]
$U_\rho \in L^{12}(\R^3) \cap W_{loc}^{2,1+1/n}(\R^3)$ with
\[
\nabla U_\rho (x) = \frac{x}{r}U_\rho'(r) = \frac{x}{r} \frac{m_\rho(r)}{r^2},
\ r = \n{x} >0,
\]
where
\[
m_\rho (r) = 4 \pi \int_0^r \rho (s)\,s^2ds = \int_{\n{x} \leq r} \rho(x)\,dx.
\] 
\item[(b)]
For every $R >0$,
\[
\int |\nabla U_\rho|^2dx
\leq
\frac{3 n}{3-n} \left(\frac{4 \pi}{3}\right)^{1+1/n} M^{1-1/n} 
R^{(3-n)/n} \int_{\n{x}\leq R} \rho^{1+1/n} dx
+ \frac{4 \pi M^2}{R}.
\]
\end{itemize}
\end{lemma}

\noindent
{\bf Proof.} As to (a) we note that $1+1/n > 4/3$ so that
$U_\rho \in L^{12}(\R^3)$ by the generalized Young's inequality.
The remaining assertions in (a) follow
from spherical symmetry. As to (b),
\beas
\int |\nabla U_\rho|^2dx
&\leq&
4 \pi \int_0^R r^{-2} m^2_\rho (r) dr + 
4 \pi \int_R^\infty r^{-2} m^2_\rho (r) dr\\
&\leq&
4 \pi M^{1-1/n} \int_0^R r^{-2} m^{1+1/n}_\rho (r) dr + \frac{4 \pi M^2}{R},
\eeas
since $m_\rho (r) \leq M,\ r \geq 0$.
By H\"older's inequality,
\beas
m_\rho (r) 
&\leq&
\left(\frac{4 \pi}{3}r^3 \right)^{1/(1+n)}
\left(\int_{\n{x} \leq R} \rho^{1+1/n}dx\right)^{n/(1+n)},\ r \leq R,
\eeas
and the assertion follows. \hfill {\it QED}

\begin{lemma} \label{lower}
Let $Q$ satisfy assumption (Q1).
Then there exists a constant $C_M$ depending on $M$ such that
\[
\D (f) \geq {1\over 2} J(f) - C_M,\ f \in \F_M,
\]
in particular,
\[
\D_M := \inf\, \{\D(f) \mid f \in \F_M\} > - \infty .
\]
\end{lemma}

\noindent
{\bf Proof.} Let $f \in \F_M$. By assumption (Q1),
\beas
\int\!\!\int f^{1+1/\mu_1}dv\,dx 
&=&
\int\!\!\int_{\{f \leq F_0\}}\ldots\;
+ \int\!\!\int_{\{f \geq F_0\}}\ldots 
\leq
F_0^{1/{\mu_1}}\int\!\!\int f + C_1^{-1} \int\!\!\int Q(f,L) \\
&\leq&
C\, (M + J(f)).
\eeas
Thus by Lemma~\ref{rhoest},
\[
\int \rho^{1+1/n}dx \leq C (M + J(f))
\]
where $1\leq n=1+1/\mu_1 < 3$. In particular, Lemma~\ref{uest} applies, and
\[
\D(f)
\geq
J(f) \left(1-C M^{1-1/n} R^{(3-n)/n}\right) - C\left(M^{2-1/n} R^{(3-n)/n}
+ \frac{M^2}{R}\right)
\]
where $C>0$ is some constant which does not depend on $R$.
In dependence of $M$ we now choose $R>0$ such that the 
term in the first parenthesis
equals $1/2$, and the proof is complete. \hfill {\it QED}

\section{Scaling and Splitting}
\setcounter{equation}{0}

The behaviour of $\D$ and $M$ under scaling transformations can be
used to relate the $\D_M$'s for different values of $M$:

\begin{lemma} \label{scaling}
Let $Q$ satisfy the assumptions (Q2) and (Q3). Then 
$-\infty < \D_M  < 0$ for each $M>0$,
and there exists $\alpha >0$ such that for all $0< M_1 \leq M_2$,  
\[
\D_{M_1}\ge \left( {{M_1}\over{M_2}}\right )
^{1+\alpha}\D_{M_2}. 
\]
\end{lemma}

\noindent
{\bf Proof.} 
Given any function $f$, we define a rescaled function 
 $\bar f(x,v)=af(bx,cv)$, where $a,\ b,\ c >0$. Then
\be \label{mscale}
\int\!\!\int \bar f\,dv\,dx = ab^{-3}c^{-3}\int\!\!\int f\,dv\,dx
\ee
and
\bea
\D(\bar f) 
&=& 
b^{-3}c^{-3}\int\!\!\int Q(af,b^{-2}c^{-2}L)\,dv\,dx \nonumber \\
&&
{}+ a b^{-3} c^{-5} \frac{1}{2} \int\!\!\int |v|^2 f\,dv\,dx
- a^2b^{-5} c^{-6}\frac{1}{8\pi}
\int |\nabla U_f|^2\,dx . \label{dscale}
\eea

\noindent
{\em Proof of $\D_M <0$:} 
Fix some $f\in \F_1$ with compact support and $f\leq F_0$.
Let 
\[
a=M b^3c^3
\]
so that
\[
\int\!\!\int \bar f \, dv\, dx = M.
\]
The last term in $\D(\bar f)$ is negative and of the order $b$,
and we want to make this term dominate the others as $b \to 0$.
Choose  $c=b^{-\gamma/2}$ and assume that $a \leq 1$ so that
$af \leq F_0$. By (Q2),
\[
\D(\bar f) \leq
C M^{1+1/{\mu_2}}b^{\frac{3}{\mu_2}(1-\gamma/2)} + C M b^\gamma -
M^2 \overline{C} b 
\]
where $C,\ \overline{C} >0$ depend on $f$. Since
we want the last term to dominate
as $b \to 0$, we need $\gamma >1$ and $3 (1-\gamma/2)/\mu_2 >1$,
and, in order that $a \leq 1$ as $b \to 0$, also
$\gamma < 2$. Such a choice of $\gamma$ is possible since $\mu_2 < 3/2$,
and thus $\D_M < 0$ for $b$ sufficiently small.

\noindent
{\em Proof of the scaling inequality if $0<\mu_3 < 1/2$:}
Assume that $f \in \F_{M_2}$ and $\bar f \in \F_{M_1}$ so that by 
(\ref{mscale}),
\be \label{m1m2}
a b^{-3}c^{-3} = \frac{M_1}{M_2} =: m \leq 1.
\ee
By (\ref{dscale}) and (Q3),
\[
\D(\bar f) \geq m a^{1/\mu_3} \int\!\!\int Q(f,L)\,dv\,dx + m c^{-2}
\frac{1}{2} \int\!\!\int \n{v} f\,dv\,dx
- m^2 b \frac{1}{8 \pi} \int \n{\nabla U_f}^2 dx
\]
provided $a \leq 1$ and $b^{-2} c^{-2} \leq 1$. Now we require that
\[
m a^{1/\mu_3} = m c^{-2} = m^2 b.
\]
Together with (\ref{m1m2}) this determines $a,\ b,\ c$ in terms of $m$.
In particular,
\[
a=m^{4 \mu_3/(3-2\mu_3)} \leq 1,\ 
b c = m^{(1-2\mu_3)/(2\mu_3-3)} \geq 1
\]
as required---recall that $0 < \mu_3 < 1/2$ in the present case---and
\[
m a^{1/\mu_3} = m^{1+\alpha},\ \alpha = 4/(3 - 2 \mu_3) 
\]
whence
\[
\D(\bar f) \geq m^{1+\alpha}\D(f).
\]
Since for any given choice of $a,\ b,\ c$ the mapping $f \mapsto \bar f$
is one-to-one and onto between $\F_{M_2}$ and $\F_{M_1}$ 
the scaling inequality follows.

\noindent
{\em Proof of the scaling inequality if $\mu_3 \geq 1/2$:} 
In this case we choose $a=b=c^{-1}$. If $f \in \F_{M_2}$ and
$\bar f \in \F_{M_1}$ then again (\ref{m1m2}) holds.
Thus $a = m \leq 1$, and since $1+ 1/\mu_3 \leq 3$,
\beas
\D(\bar f)
&\geq&
a^{1+1/\mu_3} \int\!\!\int Q(f,L)\, dv\, dx 
+ a^3 \left(\frac{1}{2} \int\!\!\int \n{v}^2 f\,dv\,dx 
- \frac{1}{8\pi} \int\n{\nabla U_f}^2 dx \right)\\
&\geq&
m^3 \D(f),
\eeas
which proves the scaling assertion in this case. \hfill {\it QED}   

We now prove a splitting estimate which is crucial to 
find a minimizer of $\D$. We define the ball $B_R
=\{x \in \R^3 \mid |x|\le R\}$.

\begin{lemma} \label{split}
Let $Q$ satisfy the assumptions (Q2) and (Q3), let $f\in \F_M$, and
\[
m_f(R) := \int_{B_R}\int f\,dv\,dx,\ R>0.
\]
Then
\[
\D (f)- \D_M
\geq
\left(-{{C_\alpha  \D_M}\over{M^2}}
-{1\over {R}}\right)\, m_f(R)\, (M-m_f(R)),\ R>0,
\]
where the constant $C_\alpha >0$ depends on $\alpha$ from Lemma~\ref{scaling}.
\end{lemma}

\noindent
{\bf Proof.} 
Let $1_{B_R\times \R^3}$ be the characteristic function of $B_R \times \R^3$,
\[
f_1=1_{B_R\times \R^3} f,\ f_2 = f - f_1
\]
and let $\rho_i$ and $U_i$ denote the induced spatial densities
and potentials respectively, $i=1,2$. We abbreviate $\lambda=M-m_f(R)$.
Then
\beas
\D (f)
&=&
J(f_1)+J(f_2)
-{1\over 8 \pi}\int|\nabla U_1|^2
-{1\over 8 \pi}\int|\nabla U_2|^2 - 
\frac{1}{4\pi}\int\nabla U_1 \cdot \nabla U_2\\
&\geq&
\D_{M-\lambda} + \D_{\lambda}
-  \frac{1}{4\pi}\int\nabla U_1 \cdot \nabla U_2.
\eeas
since $f_1 \in \F_{M-\lambda}$ and $f_2 \in \F_\lambda$.
By Lemma~\ref{uest} (a), $\nabla U_2=0$ on $B_R$, and
\[
\int\nabla U_1 \cdot \nabla U_2
\leq \lambda (M-\lambda) \,
4\pi \int_R^\infty \frac{1}{r^2}dr =
\frac{4\pi}{R} \lambda (M-\lambda).
\]
Using Lemma~\ref{scaling} we find that
\[
\D(f) \geq
\left[(1-\lambda/M)^{1+\alpha}+(\lambda/M)^{1+\alpha}\right] \D_M
- \frac{1}{R} \lambda (M-\lambda) .
\]
Since $\alpha>0$, there is $C_\alpha>0$, such that 
\[
(1-x)^{1+\alpha}+x^{1+\alpha}-1\le -C_\alpha (1-x)x,\ 0\le x\le 1. 
\]
Choosing $x=\lambda/M$ and noticing that $\D_M<0$, 
we have 
\beas
\D (f) - \D_M
&\geq&
\left[(1-\lambda/M)^{1+\alpha}
+(\lambda/M)^{1+\alpha}-1 \right] \D_M - 
\frac{1}{R} \lambda (M-\lambda)\\
&\geq&
-C_\alpha \D_M 
\left(1-{\lambda \over M}\right){\lambda\over M} 
- \frac{1}{R} \lambda (M-\lambda)\\
&=&
\left(-\frac{C_\alpha \D_M}{M^2} - \frac{1}{R}\right)\,
(M-\lambda)\lambda,
\eeas
and the proof is complete. \hfill {\it QED}

\section{Minimizers of $\D$}
\setcounter{equation}{0}

Before we show the existence of a minimizer of $\D$ over the set $\F_M$
we use Lemma~\ref{split} to show that along
a minimizing sequence the mass has to concentrate in a certain ball:

\begin{lemma} \label{r0}
Let $Q$ satisfy the assumptions (Q2) and (Q3), and define
\[
R_0:= -\frac{M^2}{C_\alpha \D_M}.
\]
If $(f_n) \subset \F_M$ is a minimizing sequence of $\D$, then
for any $R > R_0$,
\[
\lim_{n\to \infty} \int_{|x|\ge R} \int f_n dv\,dx = 0.
\]
\end{lemma}

\noindent
{\bf Proof.}   
If not, there exist some $R>R_0$, 
$\lambda>0$, and a subsequence, called $(f_n)$ again, such that 
\[
\lim_{n\to \infty}
\int_{|x|\ge R}\int f_n dv\,dx = \lambda .
\]
For every $n\in \N$ we can now choose $R_n>R$ such that
\[
\lambda_n := \int_{|x|\geq R_n}\int f_n dv\,dx = {1\over 2}
\int_{|x|\ge R}\int f_n dv\,dx .
\]
Then 
\[
\lim_{n\to \infty}
\int_{|x|\ge R_n}\int f_n dv\,dx =
\lim_{n\to\infty}
\lambda_n = \lambda/2>0.
\]
Applying Lemma~\ref{split} to $B_{R_n}$ we get 
\begin{eqnarray*}
\D (f_n) - \D_M
&\geq& 
\left(-{{C_\alpha  \D_M}\over
{M^2}}-\frac{1}{R_n}\right) (M-\lambda_n) \lambda_n
> \left(-{{C_\alpha  \D_M}\over
{M^2}}-\frac{1}{R}\right) (M-\lambda_n) \lambda_n\\
&\to&
\left(-{{C_\alpha  \D_M}\over{M^2}}-\frac{1}{R} \right)
(M-\lambda/2) \lambda/2 > 0
\end{eqnarray*}
as $n\to\infty$, since by choice of $R_0$ the expression in the
parenthesis is positive for $R>R_0$, and $0<\lambda/2<M$.
This contradicts the fact that $(f_n)$ is a minimizing 
sequence. \hfill {\it QED}
 
\begin{theorem} \label{exminim}
Let $Q$ satisfy the assumptions (Q1)--(Q4), and
let $(f_n) \subset \F_M$ be a minimizing sequence of 
$\D$. Then there is a minimizer $f_0$ and a subsequence
$(f_{n_k})$ such that 
$\D (f_0) = \D_M$, $\supp f_0 \subset B_{R_0}\times \R^3$ with 
$R_0$ as in Lemma~\ref{r0},
and $f_{n_k} \rightharpoonup f_0$ weakly in 
$L^{1+1/\mu_1} (\R^6)$.
For the induced potentials we have
$U_{n_k} \to U_0$ strongly in $L^2 (\R^3)$.
\end{theorem}

\noindent
{\bf Proof.} 
By Lemma~\ref{lower}, $(J(f_n))$ is bounded. Let
$p_1=1+1/\mu_1$. By assumption (Q1),
\[
\int\!\!\int f_n^{p_1}dv\,dx 
\leq
C \int\!\!\int f_n dv\,dx +  C \int\!\!\int Q(f_n,L)\,dv\,dx
\]
so that $(f_n)$ is bounded in $L^{p_1} (\R^6)$. 
Thus there exists a weakly convergent
subsequence, denoted by $(f_n)$ again, i.~e.,
\[
f_n \rightharpoonup f_0\ \mbox{weakly in }\ L^{p_1} (\R^6).
\]
Clearly, $f_0 \geq 0$ a.~e.,\ and $f_0$ is spherically symmetric.
Since
\beas
M 
&=& 
\lim_{n \to \infty} \int_{\n{x}\leq R_1}\int_{\n{v}\leq R_2}f_n dv\,dx
+ \lim_{n \to \infty} \int_{\n{x}\leq R_1}\int_{\n{v}\geq R_2}f_n dv\,dx\\
&\leq&
\lim_{n \to \infty} \int_{\n{x}\leq R_1}\int_{\n{v}\leq R_2}f_n dv\,dx
+ \frac{C}{R_2^2}
\eeas
where $R_1 > R_0$ and $R_2 >0$ are arbitrary, it follows that
\[
\int_{\n{x}\leq R_1}\int f_0 dv\,dx = M
\]
for every $R_1 > R_0$. This proves the assertion on $\supp f_0$
and $\int\!\!\int f_0 =M$. Also by weak convergence 
\be \label{wke}
\int\!\!\int \n{v}^2 f_0 dv\,dx \leq \liminf_{n \to \infty}
\int\!\!\int \n{v}^2 f_n dv\,dx < \infty.
\ee
By Lemma~\ref{rhoest} $(\rho_n)=(\rho_{f_n})$ is bounded in 
$L^{1+1/n_1} (\R^3)$ where
$n_1 = \mu_1 + 3/2$. After extracting a further subsequence, we thus have
that
\[
\rho_n \rightharpoonup \rho_0:=\rho_{f_0} \
\mbox{weakly in }\ L^{1+1/n_1} (\R^3) .
\]
Since $1+1/n_1 > 4/3$ 
\[
\nn{\nabla U_n -\nabla U_0}_{L^q(\R^3)} \leq C,\ n \in \N
\]
with some $q > 12/5 > 2$ by Young's inequality. On the other hand,
the compact embedding $W^{2,1+1/n_1}(B_R) \subset W^{1,1} (B_R)$
implies that 
\[
\nn{\nabla U_n -\nabla U_0}_{L^1(B_R)} \to 0 ,\ n \to \infty 
\]
for any $R>0$. By the usual interpolation argument,
\[
\nn{\nabla U_n -\nabla U_0}_{L^2(B_R)} \to 0 ,\ n \to \infty , 
\]
but since by spherical symmetry,
\[
\int |\nabla U_n-\nabla U_0|^2dx
\leq \int_{|x|\leq R}|\nabla U_n-\nabla U_0|^2 dx
+ 4 \pi \frac{M^2}{R}
\]
the convergence of the fields holds in $L^2(\R^3)$.

It remains to show that $f_0$ is actually a minimizer, in
particular, $J(f_0) < \infty$ so that $f_0 \in \F_M$.
By Mazur's Lemma there exists a sequence $(g_n) \subset L^{p_1}(\R^6)$ 
such that $g_n \to f_0$ strongly in $L^{p_1}(\R^6)$ and
$g_n$ is a convex combination of $\{f_k\mid k\geq n\}$.
In particular, $g_n \to f_0$ a.~e.\ on $\R^6$. By (Q4) the functional
\[
f \mapsto \int\!\!\int Q(f,L)\,dv\,dx
\]
is convex. Combining this with Fatou's Lemma implies that
\[
\int\!\!\int Q(f_0,L)\,dv\,dx \leq 
\liminf_{n \to \infty} \int\!\!\int Q(g_n,L)\,dv\,dx
\leq \limsup_{n\to \infty} \int\!\!\int Q(f_n,L)\,dv\,dx.
\]
Together with (\ref{wke}) this implies that
\[
J(f_0) \leq \lim_{n\to \infty} J(f_n) < \infty ;
\]
note that $\lim_{n\to \infty} J(f_n)$ exists.
Therefore,
\[
\D(f_0) = J(f_0) - \frac{1}{8 \pi} \int \n{\nabla U_0}^2 
\leq 
\lim_{n\to \infty} \left(J(f_n) - \frac{1}{8 \pi} \int\n{\nabla U_n}^2\right)
= \D_M,
\]
and the proof is complete. \hfill {\it QED}

\begin{theorem} \label{propminim}
Let $Q$ satisfy the assumptions (Q1)--(Q5), and
let $f_0 \in \F_M$ be a minimizer of $\D$. Then
\[
f_0 (x,v)=\left\{
\begin{array}{ccl} 
q(E_0-E, L)&,& E_0 - E > 0,\\
0 &,& E_0 - E\le 0
\end{array}
\right.
\]
where
\[
E = \frac{1}{2} \n{v}^2 + U_0 (x),
\]
\[
E_0 = {1\over M} \int\!\!\int \left( \partial_f Q(f_0, L)
+ E \right)\,f_0\,dv\,dx  <0,
\]
$U_0$ is the potential induced by $f_0$, and $q$ is as defined in (\ref{qdef}).
Moreover, $f_0$ is a steady state of the Vlasov-Poisson
system.
\end{theorem}

\noindent
{\bf Proof.} 
Let $f_0$ be a minimizer. We shall use the standard method of 
Euler-Lagrange multipliers to prove the theorem. 
For any fixed $\epsilon>0$ let $\eta : \R^6 \to \R$ be measurable,
with compact support, spherically symmetric, and such that
\[
\n{\eta} \leq 1,\ \  
\eta \geq 0\ \mbox{a.~e.\ on}\ \R^6 \setminus \supp f_0,\ \ 
\epsilon \leq f_0 \leq \frac{1}{\epsilon}\ 
\mbox{a.~e.\ on}\ \supp f_0 \cap \supp \eta .
\]
Below we will occasionally argue pointwise on $\R^6$ so we choose
a representative of $f_0$ satisfying the previous estimate pointwise
on $\supp f_0 \cap \supp \eta$. 
For 
\[
0 \le h \le {{\epsilon}\over{2(1+\|\eta\|_1)}}
\] 
we define 
\[
g(h)= M {{h\eta+f_0}\over {\|h\eta+f_0\|_1}}. 
\]
Clearly, 
\[
M-\frac{\epsilon}{2} \leq \|h\eta+f_0\|_1
\leq M + \frac{\epsilon}{2}.
\]
On $\supp f_0$ we have 
\be \label{ghbound}
\frac{1}{4}\,f_0 \le g(h) \le 2 f_0
\ee
provided $0<\epsilon < \epsilon_0$ with $\epsilon_0>0$ sufficiently small.
Note that $g(0)=f_0$. We expand $\D(g(h))-\D(f_0)$ in powers of $h$.
Obviously,
\bea
&&
\D(g(h))-\D(f_0) \nonumber \\
&&\qquad =
\int\!\!\int \Bigl( Q(g(h),L)-Q(f_0,L)\Bigr)\,dv\,dx + 
{1\over 2} \int\!\!\int|v|^2(g(h)-f_0)\,dv\,dx \nonumber\\
&&
\qquad\ 
{}- \frac{1}{8\pi} 
\int \left(\n{\nabla U_{g(h)}}^2 - \n{\nabla U_0}^2\right)\,dx \nonumber\\
&&\qquad =
\int\!\!\int \Bigl( Q(g(h),L)-Q(f_0,L)\Bigr)\,dv\,dx
+ {1\over 2} \int\!\!\int|v|^2(g(h)-f_0)\,dv\,dx \nonumber \\
&&\qquad\
{}+ \int\!\!\int U_0 (g(h)-f_0)\,dv\,dx - 
\frac{1}{8\pi} 
\int \n{\nabla U_{g(h)} - \nabla U_0}^2 dx . 
\label{dexpan}
\eea
Now observe that
$g(h)\ge 0$ on $\R^6$. Thus $g(h)$ is differentiable
with respect to $h$, and we write $g'(h)$ for this
derivative. Note that both $g(h)$ and $g'(h)$ are actually functions
of $(x,v)\in \R^6$, but we suppress this dependence.
We obtain
\beas
g'(h)
&=& 
\frac{M}{\|h\eta+f_0\|_1}\eta - M 
\frac{h \eta + f_0}{\|h\eta + f_0\|_1^2} \int\!\!\int\eta\,dv\,dx, \\
g''(h)
&=&
- 2\frac{M}{\|h\eta + f_0\|_1^2} \left(\int\!\!\int \eta\,dv\,dx\right)\, \eta
+2 M \frac{h\eta + f_0}{\|h\eta + f_0\|_1^3} 
\left(\int\!\!\int\eta\,dv\,dx\right)^2 .
\eeas
Now
\be \label{gprime}
g'(0) = \eta - \frac{1}{M} \left(\int\!\!\int \eta\,dv\,dx\right)\, f_0
\ee
and
\[
\n{g''(h)} \leq C\,(\n{\eta} + f_0)
\]
so that on $\R^6$,
\[
\left|g(h) - f_0 - h g'(0)\right|
\leq C\,h^2 (\n{\eta} + f_0);
\]
in the following, constants denoted by $C$ may depend on $f_0$, $\eta$,
and $\epsilon$ but never on $h$.
We can now estimate the last three terms in (\ref{dexpan}):
\bea
\int\!\!\int|v|^2(g(h)-f_0)\,dv\,dx
&=&
h \int\!\!\int \n{v}^2 g'(0)\,dv\,dx + O(h^2), \label{term3}\\
\int\!\!\int U_0 (g(h)-f_0)\,dv\,dx 
&=&
h \int\!\!\int U_0\, g'(0)\,dv\,dx  + O(h^2),\label{term4}\\  
\int \n{\nabla U_{g(h)} - \nabla U_0}^2\,dx
&=& 
\int\n{\nabla U_{g(h)-f_0}}^2 dx \nonumber \\
&\leq&
C \nn{\rho_{g(h)}-\rho_0}_{6/5}^2 \leq C h^2. \label{term5}
\eea
For the last estimate we used Young's inequality and the fact that
\[
\n{\rho_{g(h)} (x) - \rho_0(x)} \leq C h \int (\n{\eta} + f_0)(x,v)\,dv.
\]
It remains to estimate the first term in (\ref{dexpan}). 
Consider first a point $(x,v) \in \supp f_0$ with $f_0(x,v)>0$. Then
\beas
Q(g(h),L)-Q(f_0,L)
&=&
\partial_f Q(f_0,L) (g(h)-f_0) + \frac{1}{2} \partial_f^2 Q(\tau,L)
(g(h)-f_0)^2 \\
&=&
h \partial_f Q(f_0,L) g'(0) + 
h^2 \frac{1}{2}\partial_f Q(f_0,L) g''(\theta)\\
&&
{}+ \frac{1}{2} \partial_f^2 Q(\tau,L) (g(h)-f_0)^2
\eeas
where $\tau$ lies between $g(h)$ and $f_0$ and $\theta$ lies between
$0$ and $h$; both $\tau$ and $\theta$ depend on $(x,v)$.
Thus
\beas
&&
\Bigl| Q(g(h),L)-Q(f_0,L) - h \partial_f Q(f_0,L) g'(0)\Bigr|\\
&& \hspace{115pt} 
\leq
C \partial_f Q(f_0,L)\, (\n{\eta}+f_0)\, h^2
+ C \partial_f^2 Q(\tau,L)  (\n{\eta}^2 +f_0^2)\, h^2.
\eeas
By (\ref{ghbound}), $\tau$ lies between $f_0/4$ and $2f_0$,
so by iterating (Q5) a finite, $h$-independent number of times 
we find
\[
\partial_f^2 Q(\tau,L) \leq C  \partial_f^2 Q(f_0,L).
\]
By (Q3) and (Q5),
\[
\left(2^{1+1/\mu_3} -1\right) Q(f_0,L)\geq
Q(2 f_0,L) - Q(f_0,L) \geq 
\partial_f Q(f_0,L)\, f_0 + C \partial_f^2 Q(f_0,L)\, f_0^2
\]
and thus
\[
\Bigl| Q(g(h),L)-Q(f_0,L) - h \partial_f Q(f_0,L) g'(0)\Bigr|
\leq C Q(f_0,L) h^2 + C \n{\eta} h^2;
\]
here we used the continuity of $\partial_f Q$ and $\partial_f^2 Q$,
the fact that $\epsilon \leq f_0 \leq 1/\epsilon$ on  
$\supp \eta \cap \supp f_0$, 
and the fact that $L$ ranges in some
compact interval if $(x,v) \in \supp\, \eta$. The above estimate
holds for any point $(x,v) \in \supp f_0$ with $f_0(x,v)>0$.
Now consider a point $(x,v)$ with $f_0(x,v)=0$. Then 
\[
g(h)= M \frac{h\eta}{\|h\eta+f_0\|_1} \leq C \n{\eta} h
\]
so that by (Q4) and (Q2),
\beas
\Bigl| Q(g(h),L)-Q(f_0,L) - h \partial_f Q(f_0,L) g'(0)\Bigr|
= Q(g(h),L)
&\leq& 
Q(C h\n{\eta},L)\\
&\leq&
C \n{\eta}^{1+1/\mu_2} h^{1+1/\mu_2}
\eeas
for $h>0$ sufficiently small. Thus
\be \label{term1}
\int\!\!\int
\Bigl| Q(g(h),L)-Q(f_0,L) - h \partial_f Q(f_0,L) g'(0)\Bigr|\,dv\,dx
\leq C h^{1+\delta}
\ee
for some $\delta >0$.  
Combining (\ref{term3}), (\ref{term4}), (\ref{term5}),
and (\ref{term1}) with the fact that $f_0$ is a minimizer we find
\beas
0 \leq \D(g(h)) - \D(f_0)
&=&
h \int\!\!\int  
\left( \partial_f Q(f_0,L) + \frac{1}{2} \n{v}^2 + U_0 \right)
\,g'(0) \, dv\,dx\\
&& 
{} + O(h^{1+\delta})
\eeas
for all $h>0$ sufficiently small. Recalling
(\ref{gprime}) and the definitions of $E$ and $E_0$ this implies that
\[
\int\!\!\int \Bigl( \partial_f Q(f_0,L)
+ E - E_0 \Bigr)\, \eta\, dv\, dx \geq 0 .
\]
Recalling the class of admissable test functions $\eta$ 
and the fact that $\epsilon >0$ is arbitrary, provided it
is sufficiently small, we conclude that 
\[
E - E_0 \geq 0 \ \ \mbox{a.~e.\ on}\ \R^6 \setminus \supp f_0
\]
and
\[
\partial_f Q(f_0,L) +  E - E_0 = 0 \ \ \mbox{a.~e.\ on}\
\supp f_0.
\]
By definition of $q$---cf.\ (\ref{qdef})---this implies that
\[
f_0 (x,v) = q(E_0-E,L)\ \mbox{a.~e.\ on}\ \R^6.
\]
By construction,
\[
\lap U_0 = \frac{1}{r^2} (r^2 U_0')' = 4 \pi \rho_0
\]
so that $(f_0,\rho_0,U_0)$ is indeed a solution of the Vlasov-Poisson
system. Since $f_0$ has compact support and 
$\lim_{r \to \infty} U_0 (r)=0$ we conclude that $E_0<0$.
\hfill {\it QED}

We conclude this section with a brief discussion of the uniqueness
of the minimizer in $\F_M$. 
First observe that since each minimizer is spherically
symmetric, has total mass $M$, support in $B_{R_0}$, and since
$\lim_{r \to \infty} U_0(r) =0$ we have
\be \label{uinf}
U_0(r) = - \frac{M}{r},\ r \geq R_0.
\ee
Since $U_0$ solves the ODE
\[
\frac{1}{r^2} (r^2 U_0')'=4 \pi \int 
q\left(E_0-\frac{1}{2} \n{v}^2 - U_0,L\right)\, dv
\]
uniqueness would follow if $E_0$ were actually independent of the minimizer.
If $Q(f)=f^{1+1/\mu}$ with $0< \mu < 3/2$ the right hand side
of the above ODE takes the form
$c(E_0-U_0)_+^{\mu+3/2}$ with some constant $c>0$. Assume we have two 
solutions $U_i$ with corresponding $E_i$, $i=1,2$. Then in the
terminology of \cite{BP}, $\phi_i := E_i-U_i$ are E-solutions
of the Emden-Fowler equation
\[
\frac{1}{r^2} (r^2 \phi')'=- c\, (\phi)_+^{\mu+3/2}.
\]
As is shown in \cite{BP}, solutions of this ODE are turned into solutions
of an autonomous, planar system by the change of variables
\[
u(t):=- \frac{r \phi(r)^{\mu+3/2}}{\phi'(r)},\
v(t):=- \frac{r \phi'(r)}{\phi(r)},\ t = \ln r,
\]
and the E-solutions are all mapped onto the same orbit, 
called $C_3$ in \cite{BP}.
Reexpressed in terms of $U$ this implies that
\[
E_1 -U_1(r)=\gamma^{2/(\mu+1/2)}(E_2-U_2(\gamma r)),\ r>0,
\]
for some $\gamma >0$; note that a shift in $t$ corresponds to a scaling
in $r$. But then (\ref{uinf}) implies $\gamma =1$
and $U_1=U_2$. We have not been able to find an analogous argument
for $Q$'s of a more general form. 

\section{Dynamical Stability}
\setcounter{equation}{0}

We now investigate the dynamical stability of $f_0$. First we note
that for $f \in \F_M$,
\begin{equation}
\D (f)- \D (f_0)=d(f,f_0)-\frac{1}{8 \pi}
\|\nabla U_f-\nabla U_0\|^2_2.\label{d-d}
\end{equation}
where
\[
d(f,f_0) = \int\!\!\int \Bigl[Q(f,L)-Q(f_0, L)+(E-E_0)(f-f_0)\Bigr]\,dv\,dx.
\]

\begin{theorem} \label{stability}
Let $Q$ satisfy the assumptions (Q1)--(Q5) and
assume that the minimizer $f_0$ is unique in $\F_M$.
Then for all $\epsilon>0$ there is $\delta>0$ such that
for any solution $f(t)$ of the Vlasov-Poisson system
with $f(0)\in C^1_c (\R^6)\cap \F_M$,
\[
d(f(0),f_0) + \frac{1}{8\pi} \|\nabla U_{f(0)}-\nabla U_0\|_2^2 < \delta
\]
implies
\[
d(f(t),f_0) + \frac{1}{8\pi} \|\nabla U_{f(t)}-\nabla U_0\|_2^2 < \epsilon,
\ t \geq 0.
\]
\end{theorem}

\noindent
{\bf Proof.}
We first show that $d(f,f_0) \geq 0,\ f \in \F_M$.
For $E-E_0 \geq 0$ we have $f_0 = 0$, and thus
\[
Q(f,L)-Q(f_0, L)+(E-E_0)(f-f_0) = Q(f,L) \geq 0.
\]
For $E-E_0 <  0$,
\[
Q(f,L)-Q(f_0, L)+(E-E_0)(f-f_0) = 
\frac{1}{2} \partial_f^2 Q(\bar f,L) (f - f_0)^2 \geq 0
\]
provided $f>0$; here $\bar f$ is between $f$ and $f_0$. If $f=0$,
the left hand side is still nonnegative by continuity.

Now assume the assertion of the theorem were false. 
Then there exist $\epsilon_0>0$, $t_n>0$, and
$f_n(0)$ such that $f_n(0) \in \F_M$, and 
\[
d(f_n(0),f_0) + 
\frac{1}{8\pi} \|\nabla U_{f_n(0)}-\nabla U_0\|_2^2 = \frac{1}{n}
\]
but
\[
d(f_n(t_n),f_0) +
\frac{1}{8\pi} \|\nabla U_{f_n(t_n)}-\nabla U_0\|_2^2 \ge \epsilon_0>0.
\]
From (\ref{d-d}), we have 
$\lim_{n\to\infty} \D (f_n(0))=\D_M$. 
But  $\D (f)$ is invariant under the Vlasov-Poisson flow, hence
\[
\lim_{n\to\infty} 
\D (f_n(t_n))=\lim_{n\to\infty} 
\D (f_n(0))=\D_M.
\]
Thus, $(f_n(t_n)) \subset \F_M$
is a minimizing sequence of $\D$, and by Theorem~\ref{exminim} ,
we deduce that---up to a 
subsequence---$\|\nabla U_{{f_n}(t_n)}-\nabla U_0\|^2_2\to 0$. 
Again by (\ref{d-d}), 
$d(f_n(t_n),f_0)\to 0$, a contradiction.\hfill {\it QED}

\begin{corollary}
In addition to the assumption in Theorem~\ref{stability} assume that
\[
C_1 := \inf\left\{\partial_f^2Q(f,L) 
\mid 0 < f \leq C_0,\ 0 \leq L \leq C_0\right\}
>0
\]
for some constant $C_0 > \nn{f_0}_\infty + \max_{\supp f_0} L$.
If in the situation of Theorem~\ref{stability}
$f(0)\leq C_0$ and $L \leq C_0$ on $\supp f(0)$ then
\beas
&&\int\!\!\int_{\R^6\setminus\supp f_0} Q(f(t),L)\,dv\,dx
+ C_1 \int\!\!\int_{\supp f_0} \n{f(t)-f_0}^2 dv\,dx \\
&&
\hspace{220pt}
{}+\frac{1}{8 \pi} \nn{\nabla U_{f(t)} -\nabla U_0}_2^2 < 
\epsilon,\ t \geq 0.
\eeas
\end{corollary}

\noindent
{\bf Proof.} Since $L$ is constant along characteristics and
$\nn{f(t)}_\infty = \nn{f(0)}_\infty$,
\[
f(t) \leq C_0,\ L \leq C_0\ \mbox{on}\ \supp f(t),\ t \geq 0.
\]
Thus on $\supp f_0$,
\[
Q(f(t),L)-Q(f_0,L) + (E - E_0) (f(t)-f_0) \geq C_1 \n{f(t)-f_0}^2,
\]
cf.\ the argument in the proof of Theorem~\ref{stability}. \hfill {\it QED}

The assumptions of this corollary hold in particular for 
$Q(f,L)=f^{1+1/\mu}$ with $1\leq \mu < 3/2$ and for linear combinations
of such terms of the form (\ref{nopol}).

\noindent
{\bf Remark:} Assume the minimizer $f_0$ of $\D$ is not unique in
$\F_M$ and denote by $\M_M$ the set of all minimizers of 
$\D$ in $\F_M$. Then for each $\epsilon >0$ there exists $\delta >0$
such that for any solution $f(t)$ of the Vlasov-Poisson
system with $f(0) \in \F_M \cap C^1_c(\R^6)$,
\[
\inf_{f_0 \in \M_M} 
\left[d(f(0),f_0) + \frac{1}{8\pi} \|\nabla U_{f(0)}-\nabla U_0\|_2^2\right]
 < \delta
\]
implies
\[
\inf_{f_0 \in \M_M}
\left[d(f(t),f_0) + 
\frac{1}{8\pi} \|\nabla U_{f(t)}-\nabla U_0\|_2^2\right] < \epsilon,
\ t \geq 0.
\]
The proof works along the same lines as for Theorem~\ref{stability}.

\bigskip

\noindent
{\bf Acknowledgements:} The research of the first author
is supported in part by a NSF grant and a NSF Postdoc Fellowship.
The second author thanks the Department of Mathematics, Indiana
University, Bloomington, for its hospitality during the academic
year 97/98.

\end{document}